\documentclass[a4paper,10pt]{article}

\usepackage{graphicx,setspace}
\usepackage{bm}
\usepackage{amsmath,textcomp}

\def\be{\begin{equation}}
\def\ee{\end{equation}}
\def\beqn{\begin{eqnarray}}
\def\eeqn{\end{eqnarray}}
\def\eel#1{\label{#1}\end{equation}}
\def\ok{\overline k}

\begin{document}


\title{Wikipedia edition dynamics}

\author{Y. Gandica, F. Sampaio dos Aidos and J. Carvalho \\
\emph{Centre for Computational Physics, Department of Physics, }\\
\emph{University of Coimbra, 3004-516 Coimbra, Portugal} }

\date{\today}

\maketitle
\singlespacing
\begin{abstract}
A model for the probabilistic function followed in Wikipedia edition is presented and compared with simulations and real data.
It is argued that the probability to edit is proportional to the editor's number of previous editions (preferential attachment), 
to the editor's fitness and to an ageing factor. 
Using these simple ingredients, it is possible to reproduce the results obtained for Wikipedia edition dynamics
for a collection of single pages as well as the averaged results. 
Using a stochastic process framework, a recursive equation was obtained for the average of the number of editions  
per editor that seems to describe the editing behaviour in Wikipedia.
\end{abstract}

\footnotetext{$^{1}$Correspondence author. E-mail: ygandica@gmail.com}

\newpage \baselineskip1.0cm
\singlespacing
\section{Introduction}

The extensive monitoring of people's daily activities, in particular their online actions, yields a large 
amount of information which, adopting stochastic techniques, can be used to determine some of the 
probability distributions that govern social interactions. 
Statistical physics uses a probabilistic
description \cite{kardar} to obtain useful information about the general behaviour of many particle 
systems. In this way, it is possible to find some properties of macrosystems, regardless of the complex individual behaviour 
of each particle,
or, in the case of social systems, of each human being.

Human activity is very far from being a random process, in the sense of anybody being able to do anything at any time. 
Although each person has, at each time, the
opportunity to choose from different options, actually one is immersed into an intricate net of social rules, schedules, 
socially accepted manners, power and economic
constraints, etc., which end up determining the available usual options. 
The probability distribution for each type of human behaviour greatly depends on the specific human trait that is being studied.
Understanding the laws that
govern human activity is a great challenge for science as it may arguably be considered the most complex stochastic 
process in Nature. 

Wikipedia (WP) edition is one of these sources of probabilistic outcome \cite{anterior5_value_production}. 
Some effort has already been put in attempts to understand the evolution of WP as a network, with pages or topics 
as nodes and links between them as edges \cite{wikipedia_as_networks,wikipedia_as_network3,0902,buriol2006}. 
Several models have been proposed to describe the activity patterns of editors over different pages \cite{chromogram,coauthorship2}. 
For a single page, Wilkinson and Huberman proposed a simple stochastic mechanism to obtain the
probability distribution for that page's number of editions as a function of time, based on the simple rule that ``edits beget edits" \cite{assesing_the_value}. 
This case of preferential attachment belongs to the elemental process introduced by
Simon \cite{simon} in $1955$, in an early observation of universal patterns on linguistic, sociological and biological 
data.

The high quality of WP encyclopedia is the result of a collective effort by millions of volunteers,
in an apparently disorganized process of edition, acceptance and rejection, which works, in fact, as an effective 
and robust peer review procedure. 
Halfaker {\it et al.} studied some WP editor characteristics that lead to the process 
of selection of high quality contributions \cite{jury_peers}. 

In agreement with the concept of universality, in the 
statistical physics meaning, we propose in this paper a statistical approach for the probability of each 
editor to interact with the WP page, based on three principles already shown 
to be essential in human interaction. 
We use the preferential attachment mechanism to represent the strong tendency of users to improve and defend 
their previous contributions \cite{jury_peers}. 
This ownership feeling competes with the authority of users that are experts on the
page topic, which is expressed in our model by an increased value of a parameter that we associate to each editor and that is usually called fitness
\cite{understanding,bianconi_barabasi,DMPRE63}. 
Fitness may also describe the different drive that different persons have to push forward their opinions.
Finally, an ageing factor \cite{12030652v1,DMPRE62} is proposed, associating the time-dependent behaviour 
to an initial high motivation to edit, followed by a tendency to decrease the edition activity by theme completeness, personal saturation, 
blockage \cite{reputation_modelling} and/or any other possible personal reason.

The analytical calculations to describe the edition dynamics with the three above mentioned ingredients have produced a recursive equation for 
the average number of editions per editor that describes
qualitative and quantitatively the real behavior displayed by the WP edition dynamics.

This article is structured as follows: in section 2, the real data sample is presented. Section 3 discusses the choice of the ingredients used to describe real data while
section 4 explains in detail the model to represent the edition process. In section 5, the analytical treatment is developed and the conclusions are finally presented in section 6.  

\section{ Real data} 

The real data results shown in this work were obtained from the January 2010 dump of the English WP, 
containing 4.64 million pages, accessible at the WikiWarMonitor web page (http://wwm.phy.bme.hu/). 
The data sample used, a ``light dump", contains a reduced information listing of all the page edits (as the 
edition number and the editor identification). 
Only pages with more than $2000$ editions were analysed. They were divided into five ranges of $R$ (the ratio between the 
number of different editors involved in the editing of one page and the total number of editions of that page), from 0.1 to 0.6, in bins of 0.1. 
Only the first 2000 editions of each page were analysed and compared with the simulation results.

\section{Edition probability}

\begin{figure}[htp]
\scalebox{0.6}{\includegraphics{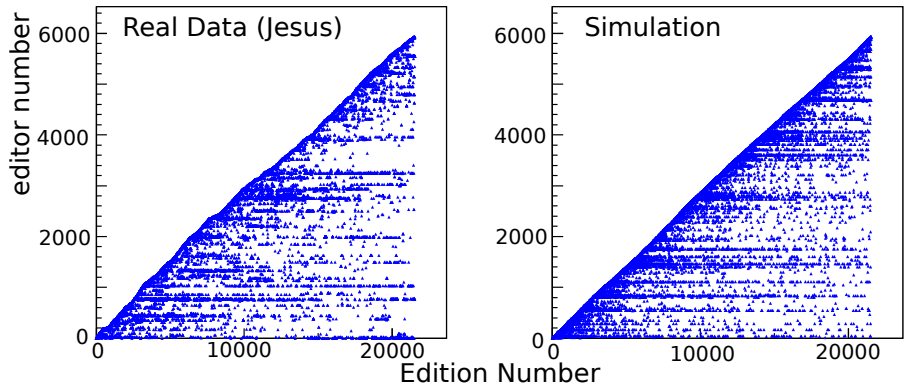}} \caption{Editor's activity as a function of edition number
for a) the WP page Jesus and b) the simulation with preferential attachment, an ageing function and 
editor's fitness, as explained in the text. In both cases, a symbol is plotted for each edition made by each editor.}
\label{fig1}
\end{figure}

It was argued in a previous paper \cite{first}, that one of the main characteristics of WP edition is an approximately constant rate, $R$, of incoming new 
editors, as illustrated in Fig. \ref{fig1}. This figure shows the editor's activity, by plotting a symbol for each edition of the article made by each 
specific editor, where editors are numbered according to the chronological order of their debut in the article. 

The real WP page Jesus (left) is compared with a simulation (right) for some chosen components and parameters, as explained later. The plot for the real 
page shows clearly an initial intensive activity for each editor, as can be seen by the high density of symbols near the diagonal (line that corresponds 
to the first edition made by each editor), followed, for most editors, by a clear decay. This indicates that most editors have an initial high motivation
to edit which, in time, just fades away. However, it is also clear from the thick long horizontal lines in the same plot, that there are some super-editors
(editors with far more editions than the average) who actually manage to maintain the editing drive.

It seems clear that we need three ingredients, to allow for a good qualitative description of this behaviour. Preferential attachment is surely an 
essential part  \cite{barabasi_albert}. 
The presence of hubs (the super-editors in WP edition) in a network is a common signature for preferential attachment. 
A fitness function is also required to enhance the edition probability of some thereafter super-editors. 
Preferential attachment alone cannot explain the 
super-editor distribution in Fig.\ref{fig1}, as all super-editors would then enter the edition process very early. 
Fitness allows for the possibility of super-editors to start at any later time. 
Finally, we must also include an ageing function to decrease the editing probability as time progresses. 
This effect is displayed in Fig.{\ref{fig1}} by the dampening of most editors' edition frequency as time elapses. 

The study of the edition process requires some definitions. 
The successive article editions are numbered in chronological order and the variable that refers to a specific edition number is denoted by $e$ where, 
in our universe of 2000 editions, $1\leq e\leq 2000$. 
The editors are also assigned a number in chronological order of their inception and the variable that refers to editor $E_i$ is $i$, where $1\leq i\leq2000\times R$. 
$e_i$ is defined as the value of $e$ when editor $E_i$ starts to edit and $\varepsilon_i=e-e_i$ is defined as the number of editions after editor $E_i$ has started to edit. 
We shall refer to an editor by $E_i$, by editor number $i$ or by editor who started at $e_i$. 
(Note that $1\leq e_i\leq 2000$ but that, for each page, there are only $2000\times R$ different values of $e_i$. Note also that $\varepsilon_i=0$ when editor $E_i$ starts to edit.)

For each range of values of $R$, all the pages with more than 2000 editions were selected. Let $N_p$ be the number of selected pages. 
We took the first 2000 editions from each of these pages and measured the number of editions done by each editor. Let $k(e_i,e)$ be the number of editions done by editor $E_i$ up to edition number $e$. 
Using $k(e_i,e)$, we made two different calculations. In one calculation, we present the results for the collection of $N_p$ pages. 
We took all the $N_p\times 2000\times R$ values of $k(e_i,e)$ and inserted them in bins with equal length (we used 1). 
Then we measured the number of editors whose number of editions laid inside each bin, thus obtaining the probability 
distribution for the number of editions per editor for that collection of pages. In the other calculation, we evaluated 
a simple average of $k(e_i,e)$ (which we denote by $\ok(e_i,e)$) for all the editors who started to edit at edition number $e_i$ over all $N_p$ different WP pages.
This is done for each of the $e$ possible values of $e_i$ (note that if, in a specific page, no editor starts to edit at edition number $e_i$, that 
counts as zero editions for the calculation, which will always average for $e$ terms). 
Finally, these $e$ values of $\ok(e_i,e)$ were inserted in bins 
of equal length (0.2) and the probability 
distribution for the average number of editions per editor was obtained.

Several simulations of the WP edition dynamics were performed with all these ingredients. 
Two different absolutely continuous random variables were tried to describe fitness: a random variable $\zeta$ with a uniform distribution in the interval 
$(0,1)$ and a random variable which follows the power law $\xi=(0.01+\zeta)^{-\gamma}$. 
We found that the uniform random variable was not powerful enough to create the 
super-editors who appear late in the edition process. The power law, on the contrary, seemed capable of providing very few editors with a very high 
edition proficiency. 
For the ageing mechanism, we tested the inactivation suggested in ref.\cite{first} and an exponential form of 
ageing $e^{-q\varepsilon_i}$, which is, for a sufficiently large pool of pages, equivalent to the inactivation procedure in the average number of 
editions per editor calculation. 
We found that this form of ageing kills the editor's contribution too quickly, thus hindering a long editorial history for the editors. 
Therefore, we tried a power law function $\varepsilon_i^{-\alpha}$, that had already been used in ref.\cite{DMPRE62}, and which
allowed for a smoother inhibition to the continued editorial activity of the editors.

The model parameters 
were chosen from a comparison between the WP real data and the simulation, using a two-sample 
Kolmogorov-Smirnov test~\cite{ks}. 
The values obtained are $\gamma=0.90$ for the fitness parameter, $q=0.0005$ for the ageing exponential parameter and $\alpha=1.25$ for the ageing power law parameter.

\begin{figure}[htp]
\scalebox{0.6}{\includegraphics{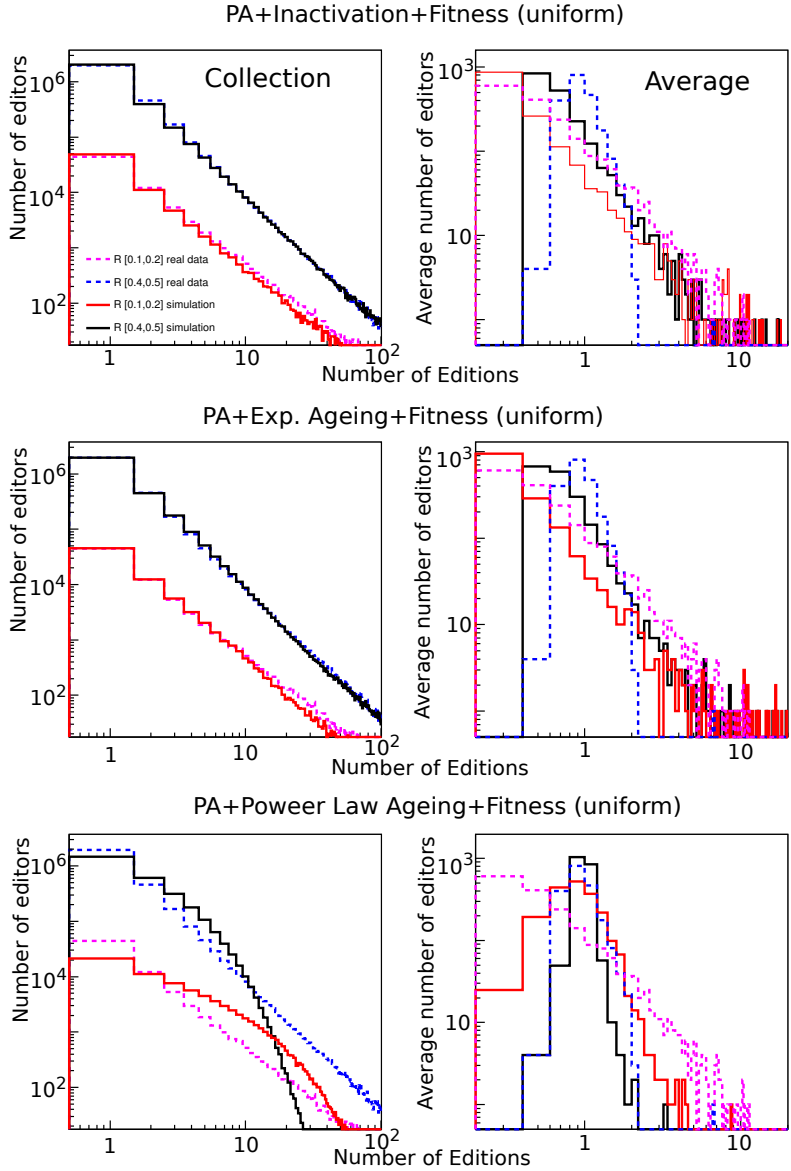}} \caption{Comparison between the number of editions per editor obtained from real data 
(dashed lines) and from simulations (full lines), for two ranges of $R$ (ratio between the number of editors and the number of editions
for each page): 
[0.1,0.2] and [0.4,0.5]. The left column shows the results for a collection of pages and the right column the results for the average over all the 
pages (the number of real and simulated WP pages is the same). 
The top row displays the results for the simulation with preferential attachment, uniform fitness and inactivation 
(see text for the explanation of the different components). 
The middle row the results for preferential attachment, uniform fitness and exponential ageing,  
and the bottom row the results for preferential attachment, uniform fitness and power law ageing.
The agreement is very poor for all combinations.}
\label{fig2}
\end{figure}

\begin{figure}[htp]
\scalebox{0.6}{\includegraphics{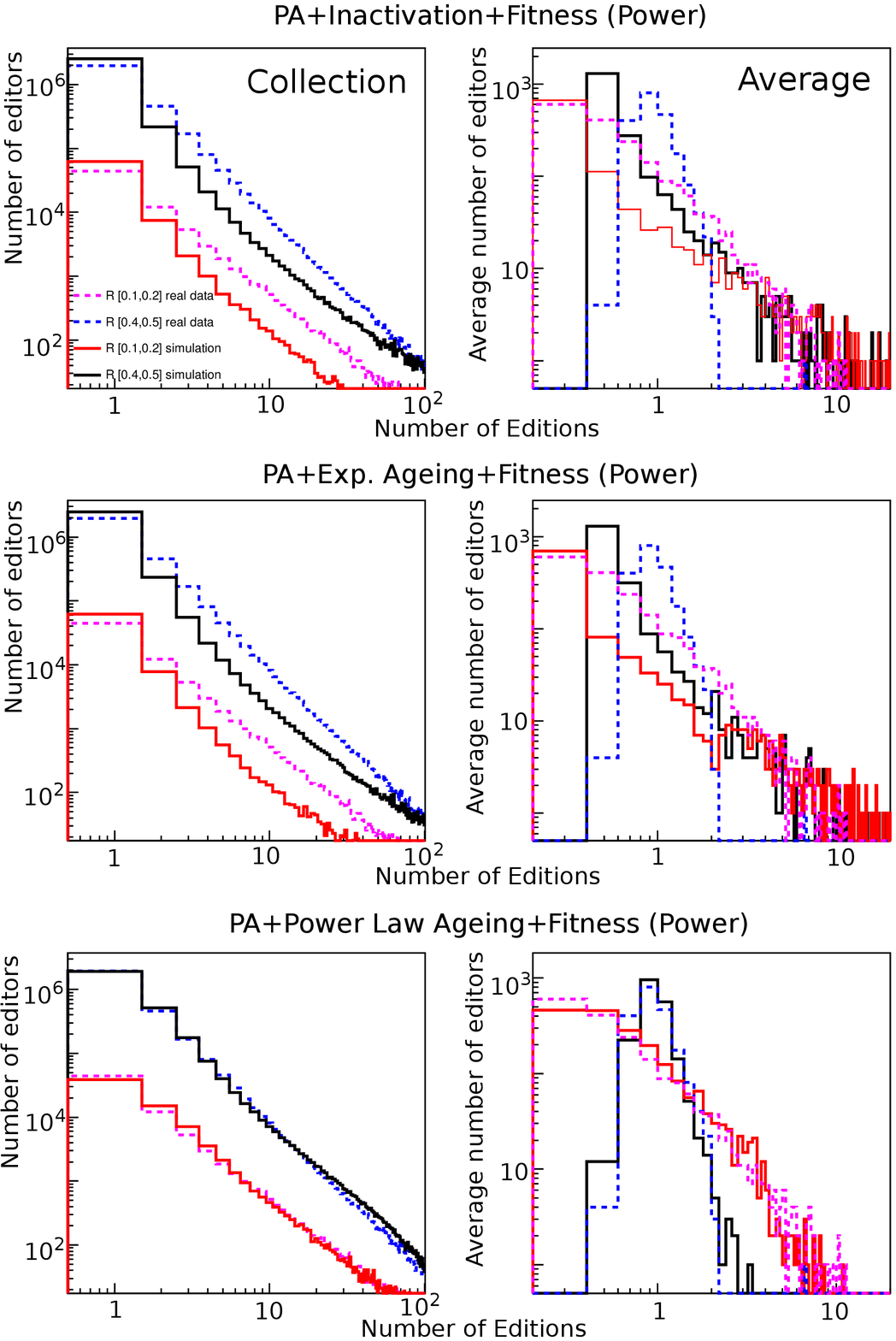}} \caption{Comparison between the number of editions per editor obtained from real data (dashed lines) and 
from simulations (full lines), for two ranges of $R$ (ratio between the number of editors and the number of editions in each page): [0.1,0.2] and [0.4,0.5].
The left column shows the results for a collection of pages and the right column the results for the average over all the pages (the number of real and  
simulated WP pages is the same). 
The top row displays the results for the simulation with preferential attachment, power law fitness and inactivation 
(see text for the explanation of the different components). 
The middle row the results for preferential attachment, power law fitness and exponential ageing,  
and the bottom row the results for preferential attachment, power law fitness and power law ageing.
Only the last combination shows a good agreement between real data and simulation for all the distributions.}
\label{fig3}
\end{figure}

In Figs. \ref{fig2} and \ref{fig3}, we show a comparison between real data and simulations. 
The same number of pages was used in simulations and in real data calculations. 
For two different ranges of $R$, we show both the number of editions per editor for a collection of pages and the average number of editions per editor (as explained above). 
In Fig. \ref{fig2}, the simulations were done with uniform fitness and each of the three kinds of ageing, and in Fig. \ref{fig3} 
the simulations were run with power law fitness and again each of the three kinds of ageing. 
The results indicate that the best fit is obtained with the power law fitness and the power law ageing mechanism.

The exponential and the power law forms of ageing were further compared in Fig. \ref{4}, which shows the average number of editions $\ok(e_i,e)$ as a function 
of $e_i$ for $e=2000$, for real data and for two simulations. Both simulations were preformed with preferential attachment and the power law form of fitness. One uses the
exponential ageing mechanism and the other uses the power law. The comparison is striking.

\begin{figure}[htp]
\scalebox{0.5}{\includegraphics{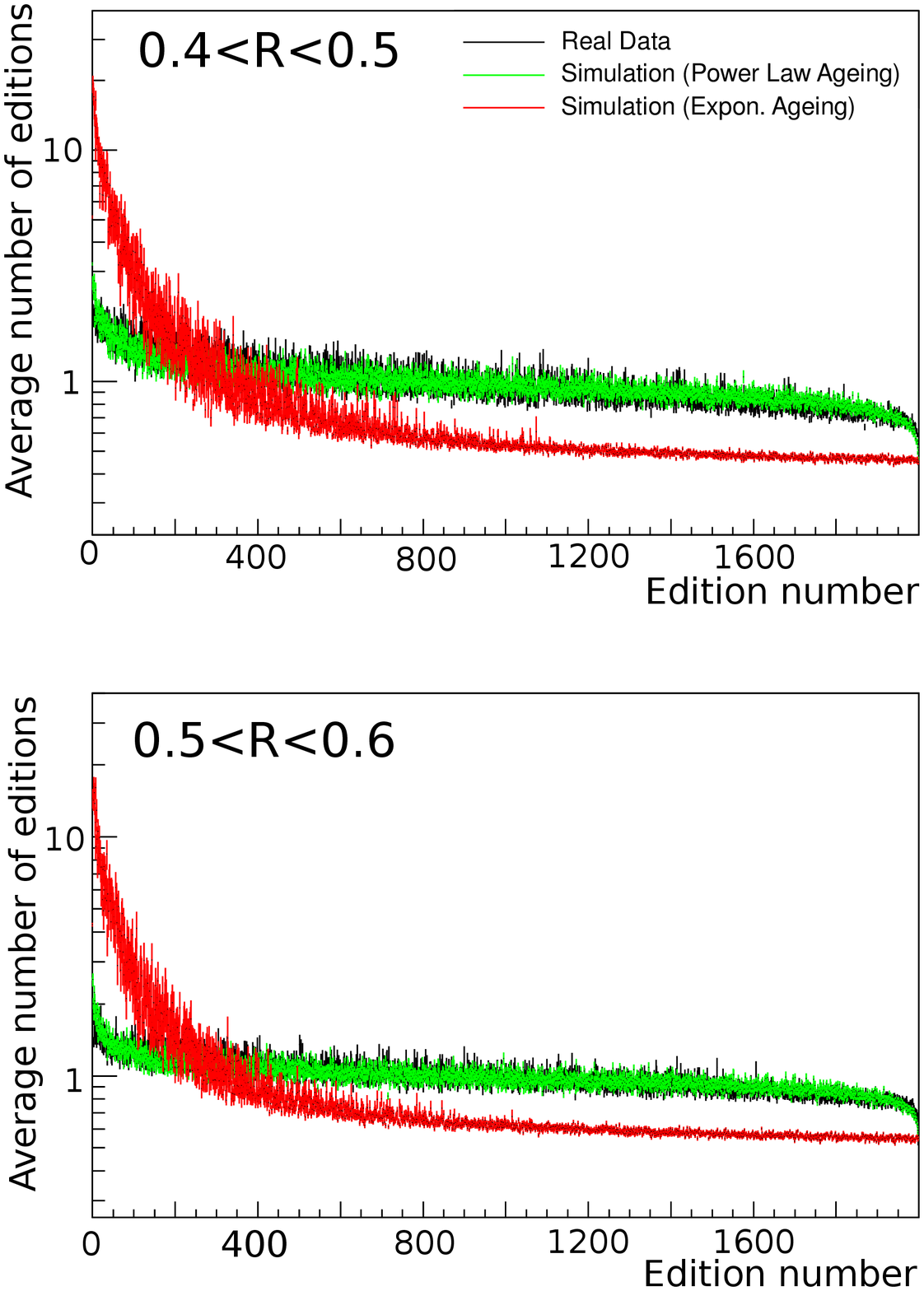}} \caption{Average number of editions per editor $\ok(e_i,e)$ as a function of $e_i$ for $e=2000$
for two ranges of $R$ (ratio between the number of editors and the number of editions in each page): top [0.4,0.5] and bottom [0.5,0.6].
The real data is shown in black, the simulation with preferential attachment, power law fitness and exponential ageing in red, and the 
simulation with preferential attachment, power law fitness and power law ageing in green. It is clear the general good description 
of real data and the later simulation.}
\label{fig4}
\end{figure}

\section{Fitting model}

The agent-based model used to obtain the simulations in this paper is based on the ingredients discussed in the previous section. 
The model is described in detail in \cite{first} but the only part that is required here are the elements on the edition dynamics.

A computer simulation run starts with one editor and the choice of a value of $R$. At each dynamical step, $e$, a new editor comes in and edits the page for the first time with probability 
$R$. Then, at each time step, an old editor $E_i$ will edit the article with probability $1-R$. The 
probability for choosing a particular editor $E_i$ among all the old editors will be:
\begin{equation}
\Pi_i= \frac{k(e_i,e-1) x_i \varepsilon_i^{-\alpha}}{\sum_j k(e_j,e-1) x_j \varepsilon_j^{-\alpha}}
\label{eq1}\end{equation}
where $x_j$ is the fitness parameter of the editor who started at $e_j$, which is initially chosen following the power law mentioned above, and is
maintained during the whole run. The sum in the denominator is over all editors who have edited the article before edition $e$.

\begin{figure}[htp]
\scalebox{0.6}{\includegraphics{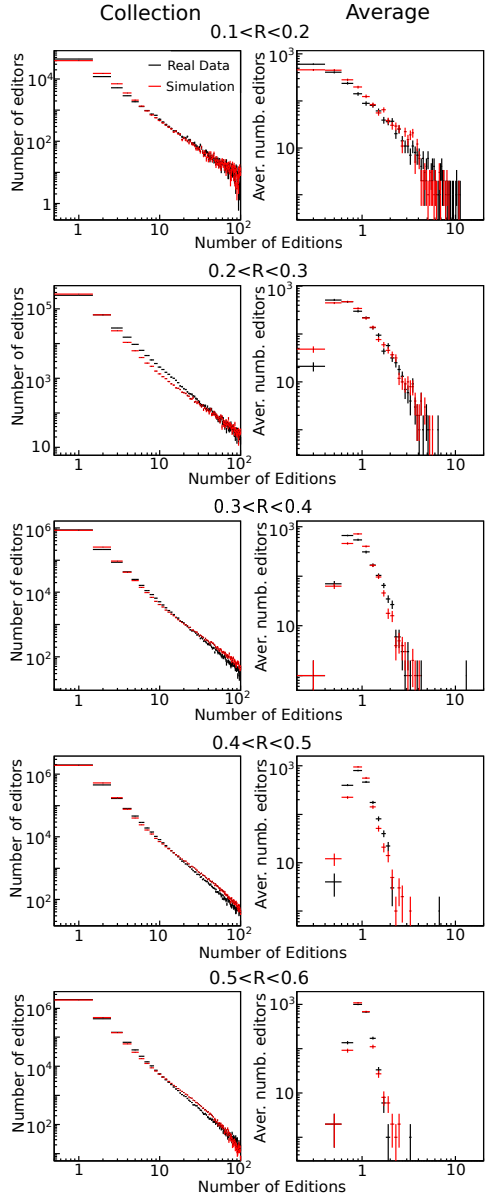}} \caption{Comparison between the number of editions per editor obtained from real data (black line) and  
simulation (red line), for five ranges of $R$ (ratio between the number of editors and the number of editions in each page), one in each row.
The left column shows the results for a collection of pages and the right column the results for the average over all the pages (the number of real and of 
simulated WP pages is the same). 
The simulation was obtained with preferential attachment, power law fitness (with $\gamma=0.90$) and power law ageing (with $\alpha=1.25$)
(see text for the explanation of the different components). 
The agreement between real data and simulation for all the distributions is, in general, good.}
\label{fig5}
\end{figure}

The qualitative agreement between real data and simulation is good for all the ranges of $R$, as shown 
in~\ref{fig5}, taking into account the reduced number of ingredients and parameters used. 

\section{Analytic calculation} 

In this section, an analytical approach for the problem of finding the average number of editions per 
editor is developed. The mathematical problem of obtaining the number of editions of an editor at each edition number $e$ is
similar to the problem of finding the degree of a node in a network. It has been mentioned before that the editors play the role of nodes 
and the editions the role of connections, the difference being, of course, that in this case the edition 
does not connect an editor to another one. Instead, it connects an editor to the WP page that is being 
edited, which is outside the network.

This problem is often studied under the continuum approximation \cite{barabasi_albert,bianconi_barabasi,
DMPRE63,DMPRE62,KRL,BAJ} although some authors have performed exact calculations \cite{KK,DMS}. A network with preferential attachment and a power law ageing has already been studied by Dorogovtsev and Mendes \cite{DMPRE62}. Using a continuum approach, they have obtained an equation for $\ok(e_i,e)$ (in the following discussion, their notation will be adapted to this paper's specific problem):
\be
	\frac{\partial \ok(e_i,e)}{\partial e}=\frac{\ok(e_i,e)(e-e_i)^{-\alpha}}{\int_0^e\,du\,\ok(u,e)(e-u)^{-\alpha}}\qquad , \qquad \ok(e,e)=1
\eel{2}
In the continuum approach it is assumed that $e$, $e_i$ and $k$ are continuous real variables. They proceeded to show that, for $\alpha<1$, the solution can be written in terms of a hypergeometric function and that the distribution of the number of editions per editor follows asymptotically a power law. For $\alpha\ge 1$, however, the problem is more complicated. It is easy to see that Eq.(\ref{2}) fails for this range of values of $\alpha$, due to a divergence of the integral in the denominator. According to Dorogovstsev and Mendes, the number of editions per editor should now follow an exponential behaviour, but the statistics is too poor for this statement to be verified by simulations.

The power law ageing effect described in this paper is the same as the one Dorogovtsev and Mendes use in the above formalism.  However, as the value obtained for the parameter $\alpha$ is larger that one, their results cannot be used. Clearly, the divergence of the integral in Eq.(\ref{2}) is caused by the continuum approach, as the divergent term $(e-u)^{-\alpha}$ would never be larger that one in the discrete approach for $\alpha\ge 1$.

Therefore it is necessary to go back to the discrete formalism. Assuming an ensemble of similar articles, the average number of editions $\ok(e_i,e)$ made by the editors who started at $e_i$ will now be
\be
	\ok(e_i,e)=\lim_{N\rightarrow\infty}\sum_{j=1}^{N}\frac{k_j(e,e_i)}{N}
\eel{3}
where $k_j(e,e_i)$ is the value of $k(e_i,e)$ for article $j$ of the ensemble (it can be zero, if no editor started at $e_i$ in that particular article). Now, let $\kappa(e,e_i)$ and $\xi(e_i)$ be the random variables ``number of editions of the editor who started at $e_i$ at edition number $e$" and ``fitness of the editor who started at $e_i$" respectively. Then
\be
	\ok(e_i,e)=\sum_{k=0}^{e-e_i+1}\,k\,P\{\kappa(e,e_i)=k\}\qquad 1\le e_i\le e
\eel{4}
where $P$ is the probability function. Naturally, the term $k=0$ does not contribute to the summation in Eq.(\ref{4}). However, it was left there to stress the non-zero probability for the editor who started at $e_i$ not to edit a specific WP page ($P\{\kappa(e,e_i)=0\}\ne 0$ for $i>1$). For $e_i=e$, we obviously have
\be
	\ok(1,1)=1\qquad \mbox{and} \qquad \ok(e,e)=R, \quad\forall_{e>1}.
\eel{4a}

The change of $\ok(e_i,e)$ in one step, from $e$ to $e+1$, is given 
by
\be
	\Delta\ok(e_i,e+1)=\overline{k(e_i,e+1)-k(e_i,e)}=\sum_{k=0}^{1}\,k\,P\{\kappa(e+1,e_i)-\kappa(e,e_i)=k\},\; 1\le e_i\le e
\eel{5}
and
\be
	\Delta\ok(e+1,e+1)=R.
\eel{eq5}

$\Pi_i$ has already been defined by Eq.(\ref{eq1}) as the probability for choosing editor $E_i$ to edit the article at edition number $e$, 
assuming that, at $e$, no new editor comes in. In this case, $\Pi_i$ is a random variable that, besides depending on $e$ and $e_i$, 
is proportional to the editor's number of editions $\kappa(e,e_i)$ and to its fitness $\xi(e_i)$. 
However, this implies that $\Pi_i$ must also depend on the number of editions, edition number and fitness of all the other editors (because of the normalisation constant for the 
probability) and the simplification of Eq.(\ref{5}) requires some care.

We define the random vectors $\bm{\kappa}(e)$ and $\bm{\xi}(e)$ and the integer and real vectors $\bm{k}(e)$ and $\bm{x}(e)$ respectively as vectors with $e$ components given by \be \left[\bm{\kappa}(e)\right]_i=\kappa(e,e_i)\qquad
	\left[\bm{\xi}(e)\right]_i=\xi(e_i)\qquad
	\left[\bm{k}(e)\right]_i=k_i\qquad
	\left[\bm{x}(e)\right]_i=x_i
\eel{eq6}
for $i=1,...,e$, where $k_i$ is a variable that can take integer values from zero up to $e$ and $x_i$ is a variable that takes values in the fitness set of values. 
$\Pi_i(\bm{k}(e),\bm{x}(e))$  (we shall omit the dependence on $e$ and $e_\ell$, for $1\leq\ell< e$) is then related to the conditional probability
\be
	P\{\kappa(e+1,e_i)-\kappa(e,e_i)=1|\bm{\kappa}(e)=\bm{k}(e),\bm{\xi}(e)=\bm{x}(e)\}=(1-R)\Pi_i(\bm{k}(e),\bm{x}(e))
\eel{10}

In Eq.(\ref{5}) only the term $k=1$ survives and we can write
\beqn
	\Delta\ok(e_i,e+1)&=&\sum_{\bm{k}(e)}\sum_{\bm{x}(e)}\,P\{\kappa(e+1,e_i)-\kappa(e,e_i)=1,\bm{\kappa}(e)=\bm{k}(e),\bm{\xi}(e)=\bm{x}(e)\}\\
	&=&(1-R)\sum_{\bm{k}(e)}\sum_{\bm{x}(e)}\,\Pi_i(\bm{k}(e),\bm{x}(e))\,P\{\bm{\kappa}(e)=\bm{k}(e),\bm{\xi}(e)=\bm{x}(e)\}\label{12}
\eeqn
where the sum over $\bm{k}(e)$ runs over all sets of integers for which there is a positive contribution to the sum. The sum over $\bm{x}(e)$ will be an integral if the fitness turns out to be a continuous variable (as is the case in our model). In order to continue, a specific expression for $\Pi_i$ must be chosen. Eq.(\ref{eq1}) yields
\be
	\Pi_i(\bm{k}(e),\bm{x}(e))=\frac{k_i\,x_i\,(e+1-e_i)^{-\alpha}}{\sum_{\ell=1}^{e} k_\ell\,x_\ell\,(e+1-\ell)^{-\alpha}}.
\eel{13}
Unfortunately, this expression does not allow for the simplification of Eq.(\ref{12}). One way to solve this problem is to make the approximation that the denominator is approximately a function of $e$ alone
\be
	F(e+1)\simeq\sum_{\ell=1}^{e} k_\ell\,x_\ell\,(e+1-\ell)^{-\alpha}.
\eel{14}

This approximation amounts to assuming that the probability distribution
$P\{\bm{\kappa}(e)=\bm{k}(e),\bm{\xi}(e)=\bm{x}(e)\}$
will cause a tight spread of the summation in eq.(\ref{14}) around an $e$ dependent value. 
Using this expression in Eq.(\ref{12}) we obtain
\be
	\Delta\ok(e_i,e+1)=\frac{1}{F(e+1)}(1-R)\,(e+1-e_i)^{-\alpha}\,\sum_{k_i,x_i}\,k_i\,x_i\,P\{\kappa(e,e_i)=k_i,\xi_i=x_i\}.
\eel{15}

As the random variables $\kappa(e,e_i)$ and $\xi_i$ are not independent, we cannot simplify this equation
by factorising the probability in two terms (the probability for $\kappa$ and the probability for $\xi$) and eq.(\ref{15}) becomes 
\be
	\Delta\ok(e_i,e+1)=\frac{1}{F(e+1)} (1-R) (e+1-e_i)^{-\alpha} E(\xi_i E(\kappa(e,e_i)|\xi_i))
\eel{16}
where $E(\kappa(e,e_i)|\xi_i)$ is the conditional 
expectation value for $\kappa(e,e_i)$, assuming $\xi_i$.

However, this expression suggests that we can obtain a more tractable result if we start with the conditional average value $\ok(e_i,e|x_i)$, 
assuming a specific value $x_i$ for the fitness of editor $E_i$, instead of starting with the non-conditional average. 
Then we do not get the final sum in $x_i$ and the sum in $k_i$ just produces the conditional average again. 
Similar calculations to the ones above lead to the following result, after using approximation (\ref{14}),
\be
	\ok(e_i,e+1|x_i)=\ok(e_i,e|x_i)+\frac{1}{F(e+1)} \,(1-R) \, (e+1-e_i)^{-\alpha} \, x_i \, \ok(e_i,e|x_i).
\eel{17}

This equation, together with Eq.(\ref{4a})
can be solved by recurrence and the result for the non-conditional average can then be obtained by
\be
	\ok(e_i,e)=\sum_{x_i}\, \ok(e_i,e|x_i)\, P\{\xi_i=x_i\}.
\eel{19}

$F(e+1)$ can be evaluated by summing Eq.(\ref{17}) for all values of $e_i$ from $1$ to $e$ and using Eqs.(\ref{4a}) and (\ref{19}). The final result is
\be
	F(e+1)=\sum_{e_i=1}^e\sum_{x_i} \, (e+1-e_i)^{-\alpha} \, x_i \, \ok(e_i,e|x_i)\, P\{\xi_i=x_i\}.
\eel{20}

We choose the values of $x_i$ that allow for the calculation of the integral in Eq.(\ref{19}) by a Gauss-Legendre quadrature.
Eq.(\ref{17}) is then solved recursively for those values of $x_i$, thus providing a value for the average number of editions done by the editor 
who starts to edit at step $e_i$, and who has fitness $x_i$ (in the approximation Eq.(\ref{20})). 
Finally, Eq.(\ref{19}) provides the results for $\ok(e_i,e)$.

\begin{figure}[htp]
\scalebox{0.6}{\includegraphics{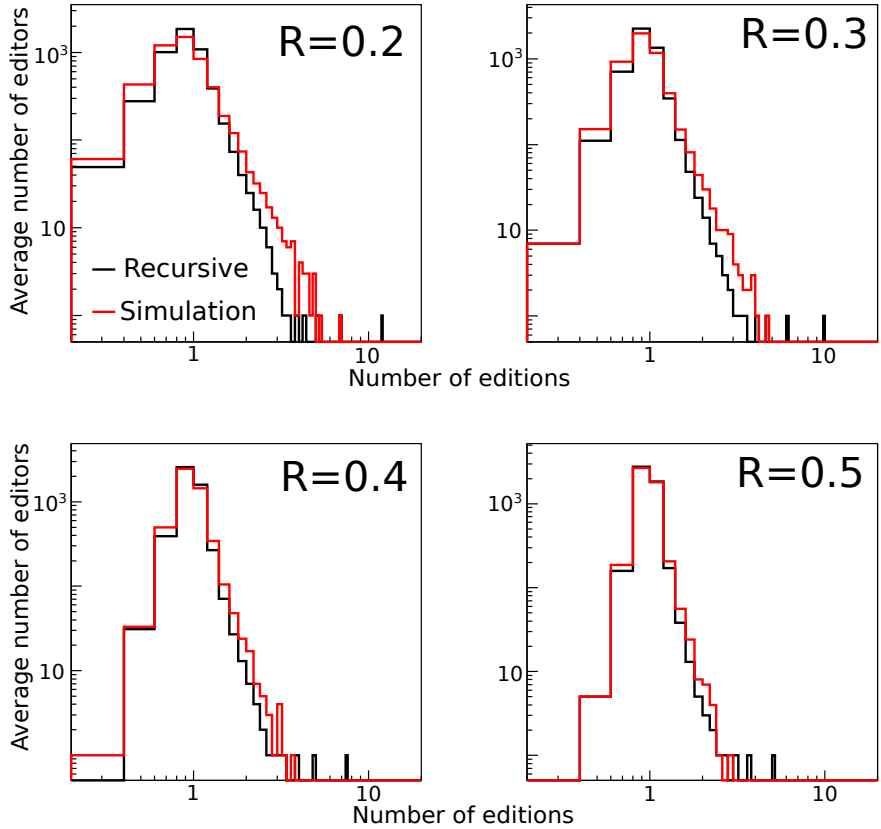}} \caption{Comparison between the average number of editions per editor obtained with the recursive approach 
(black line) and simulation (red line), for four values of $R$ (ratio between the number of editors and the number of editions 
in each page, 0.2, 0.3, 0.4 and 0.5). 
The simulation was obtained with the same ingredients as before: preferential attachment, power law fitness (with $\gamma=0.90$) and power law ageing (with $\alpha=1.25$) 
(see text for the explanation of the different components). The agreement between the recursive calculation and simulation is, in general, good.}
\label{fig6}
\end{figure}

Several number of points were tried for the Gauss-Legendre quadrature. It was found that 40 points were enough to obtain convergence, as the results almost did not change with a larger number (we went up to 100 points). The recursive calculation is compared with the simulation in Fig.~\ref{fig6}. Both calculations were performed with 5k editions and the simulation was obtained with an average over 50k pages.
The agreement is, in general, good for all the values of $R$. The discrepancies are not due to statistical error, but to the approximation in Eq.(\ref{14}). This approximation seems to be reasonable.

\section{Summary}
In this paper, Wikipedia edition, which is an available free source of human knowledge, was analysed. It was shown that this process, far from being random, is instead well reproduced in terms of previous activity, capability to edit and an ageing effect. We successfully reproduced the distribution of both the number of editions per editor for a collection of single pages and its average  showed by real data. It is proposed that the essential ingredients for the probabilistic function are
the preferential attachment, a power law ageing function and an editor random fitness, also following a power law distribution. 
The comparison of real data with the results obtained by simulations with different ingredients, 
previously found as emergent from human interactions, were shown. 
An agent-based model was developed, using the best fitting  
ingredients and it successfully reproduces real data, both for the edition of a collection of single pages and for the average over many pages. 
A recursive expression of the probabilistic function was achieved for the average WP page edition.
The agreement of the analytical approach with the simulation results is good, being also in accordance with real data.

The work of Y.G. presents research results of the Belgian Network DYSCO (Dynamical Systems, Control, and Optimization), funded by the Interuniversity 
Attraction Poles Programme, initiated by the Belgian State, Science Policy Office. The scientific responsibility rests with its author(s).

\end{document}